\documentclass[notitlepage,
aps,
pra,
showpacs,superscriptaddress,groupedaddress,
 reprint, amsmath,amssymb,onecolumn,
]{revtex4-1}

\usepackage{bm}
\usepackage{subfigure}
\usepackage{calrsfs}    
\usepackage{color}
\def\white{\textcolor{white}}

\definecolor{darkblue}{RGB}{83,0,93}
\usepackage{graphicx} %eps figures can be used instead
\usepackage{mathrsfs}

\begin{document}

\title{Propulsion and Mixing Generated by the Digitized Gait of \textit{Caenorhabditis elegans}}
\author{Ahmad Zareei}
\author{Mir Abbas Jalali}
\email{mirabbas.jalali@gmail.com}
\author{Mohsen Saadat}
\author{Peter Grenfell}
\author{Mohammad-Reza Alam}
%\email{reza.alam@berkeley.edu}
\affiliation{
Department of Mechanical Engineering, University of California, Berkeley, California 94720, USA}

%\date{\today}

\begin{abstract}
Nematodes have evolved to swim in highly viscous
environments. Artificial mechanisms that mimic the locomotory
functions of nematodes can be efficient viscous pumps. We
experimentally simulate the motion of the head segment of {\it
  Caenorhabditis elegans\/} by introducing a reciprocating and rocking
blade. We show that the bio-inspired blade's motion not only induces a flow structure similar to that of the worm, but also mixes the surrounding fluid by generating a circulatory flow. When confined between two parallel walls, the blade causes a steady Poiseuille flow through closed circuits. The pumping efficiency is comparable with the swimming efficiency of the worm. If implanted in a sealed chamber and actuated remotely, the blade can provide pumping and mixing functions for microprocessors cooled by polymeric flows and microfluidic devices.
\end{abstract}

%\pacs {45.40.-f,05.45.-a,05.10.-a}

\maketitle

%%%%%%%%%%%%%%%%%%%%
\section{Introduction}

Emerging technologies in microprocessor cooling, polymeric flows, and
in the pharmaceutical and chemical industries require the resolution
of numerous theoretical and practical challenges posed by pumping and
mixing in low Reynolds number conditions. In practice, pumping and
mixing are regarded as two different functions, and engineered systems
usually have separate units to perform these tasks. Widely used
syringe, peristaltic, and piezo pumps do not mix flowing liquids, and
few mixing protocols are known. In microchannels where the Reynolds
number is low, as the liquid is circulated by an external pump,
passive mixing is achieved by special carvings on the channel walls
\cite{stroock2002chaotic}. The well-known active mixing protocol is a
rod moving in a figure-eight trajectory \cite{gouillart2007walls}, but
this protocol has not been tested for pumping. On the other hand, in
natural systems, we know little about the interconnection between the
pumping and possible mixing of blood in small-scale living beings. For
instance, early cardiac function in embryonic vertebrates occurs in
low Reynolds number conditions, and the propagation of elastic waves
seems to pump the blood \cite{forouhar2006embryonic}. We do not know
if such traveling waves, preferred by nature, can also mix the
blood. The combination of the pumping and mixing functions in a single
artificial mechanism is an unsolved problem, which we tackle in this
work.

\begin{figure}[!h]
  \parbox[c]{.48\textwidth}{
      \mbox{\includegraphics[width=0.5\textwidth]{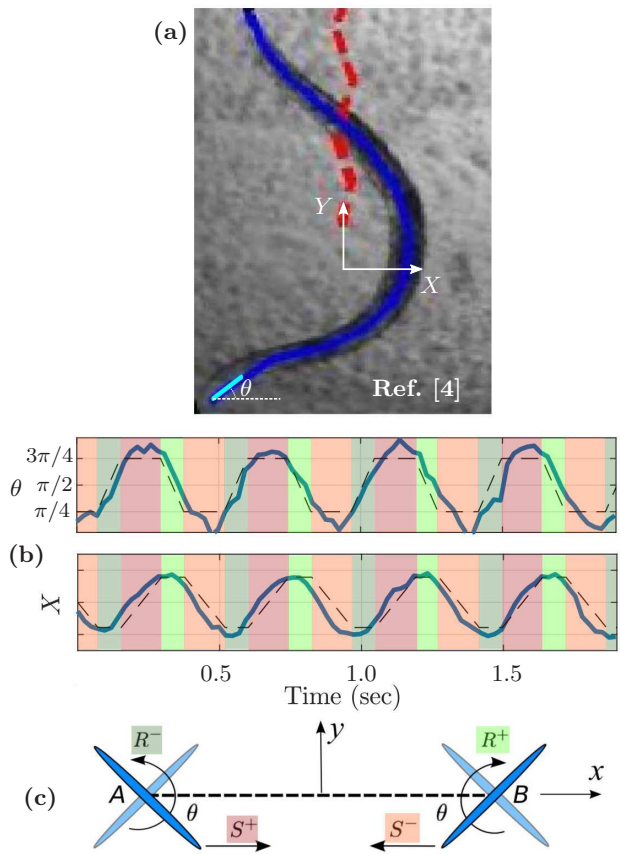}
      \put(-112,185){\white{\textbf{Ref.~\cite{{sznitman2010material}}}}}
      \put(-162,185){\white{$\theta$}}
      \put(-94,224){\white{$X$}}
      \put(-134,254){\white{$Y$}}
      \put(-239,149){\footnotesize ${\pi}/{2}$}
      \put(-243,160){\footnotesize ${3\pi}/{4}$}
      \put(-240,139){\footnotesize ${\pi}/{4}$}
      \put(-249,147){$\theta$}
      \put(-177,77){$0.5$}
      \put(-122,77){$1.0$}
      \put(-67,77){$1.5$}
    \put(-195,320){{\textbf{(a)}}}
    \put(-250,122){{\textbf{(b)}}}
    \put(-244,30){{\textbf{(c)}}}
  }}\hfill \parbox[c]{.48\linewidth}{
  \caption{Digitizing the head movement of {\it C. elegans\/}. (a)
    The shape of \textit{C. elegans} in its undulatory gait
    (cf. \cite{sznitman2010material,gray1964locomotion,stephens2008dimensionality}). The
    {\it C. elegans\/} is shown in blue, the head is shown in cyan,
    and the red dashed line represents the recorded motion of its center
    of mass. (b) The angle and position of the head of {\it C. elegans\/} in
    solid blue lines. The head pushes the surrounding fluid, switches
    its orientation after a transitional phase, then pushes the fluid
    in the opposite direction. The dashed lines represent the digitized
    approximation of the motion of head. Snapshots are taken from videos of
    the behavior of {\it C. elegans\/} in Ref. \cite{sznitman2010material}.
    (c) A schematic demonstration of the reciprocating and rocking
    motion of a blade that simulates the head of
    \textit{C. elegans}. Each stage of motion is color coded to match
    with (b).}
\label{fig1}
}
\end{figure}

A naturally inspired way of propulsion and mixing in low Reynolds
number conditions is to replicate the locomotory function of
microorganisms
\cite{dreyfus2005microscopic,shields2010biomimetic,jalali2015microswimmer,
  mirzakhanloo2018hydrodynamic,mirzakhanloo2018flow}, especially ones
that can swim in various environmental conditions by adapting their
gaits. Experimental studies show that when {\it C. elegans\/} swims in
water, the Reynolds number $Re$ exceeds 0.2
\cite{backholm2015effects}, and the worm's head undergoes a flapping
oscillation with one degree of freedom and almost no traveling wave
along its body (see Fig. \ref{fig2}A in
Ref. \cite{fang2010biomechanical}). Increasing the viscosity of the
surrounding fluid dramatically changes the gait: a quasi-periodic wave
travels from the head to tail, the undulatory wavelength drops, and
the head segment exhibits a combination of flapping and translational
motions \cite{fang2010biomechanical,sznitman2010material}. We have
analyzed videos of {\it C. elegans\/} movements and extracted a
sequence of snapshots of its body configuration over a period of
undulatory gait (Fig. \ref{fig1}A). It is seen that the head tends to
push the surrounding fluid before it flips and there exists a phase
difference between its translational and rotational motion
(Fig. \ref{fig1}B).  We ``digitize'' the movement of the head by
  approximating it with a linear translation followed by a rotational
  motion with $\pi/2$ phase difference, and model the head by a thin
  rigid blade (Fig. \ref{fig1}C). The phase difference between
translational and rotational motion breaks the time symmetry of
motion, and plays an essential role in creating a net fluid flow at
low Reynolds.
% Since the angle of attack remains at about $45^{\circ}$ over one
% complete period \cite{fang2010biomechanical}, the head generates a net thrust in the
% direction perpendicular to its lateral translation.
It is to be noted that \textit{C. elegans} gains a total thrust from
its tail, body, and head segments and here we only digitize the head
movement as a bio-inspired motion generating propulsion and mixing.

% \begin{figure*}
% \centerline{\mbox{\includegraphics[width=0.6\textwidth]{fig2.eps}}  }
% \caption{Propulsion and mixing by a reciprocating and rocking
%   blade. The blade's shaft moves along the black slit. Two rows of
%   dyes have been injected into the surface of the fluid. The lower set
%   is composed of a sequence of blue and red dyes, which have had an
%   initial distance of $4.4$ cm from the blade's path. Upper blue dyes
%   had an initial distance of $5$ cm from the slit. Snapshots have been
%   taken when $\dot x>0$ and the blade is half way from the end points
%   of its path. The stroke length is $AB=4.5$ cm. The resolution of
%   camera is $1920\times 1080$ pixels.  We have maximized the
%     exposure with linear Gamma correction of the pictures to
%   eliminate the color of the background fluid.}
% \label{fig2}
% \end{figure*}

\begin{figure*}
  \parbox[c]{.48\linewidth}{\mbox{\includegraphics[width=0.6\textwidth]{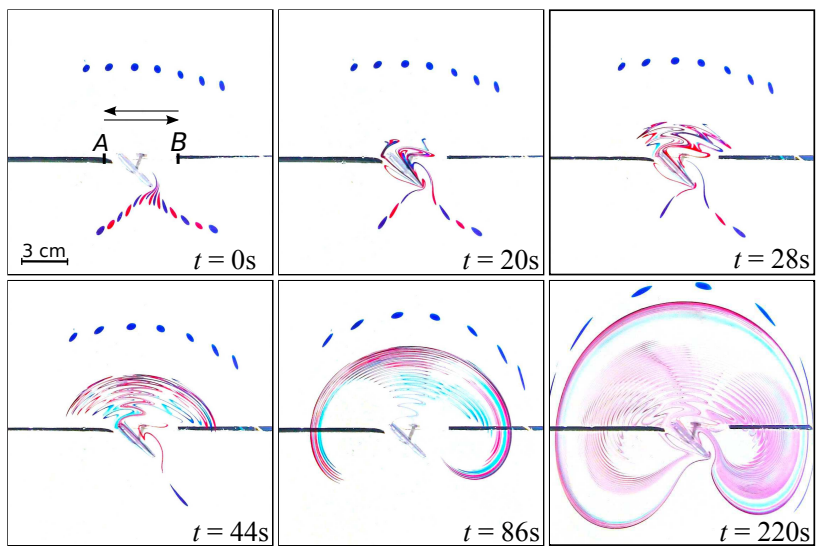}}
  }\hfill \parbox[c]{.38\linewidth}{
    \caption{Propulsion and mixing by a reciprocating and rocking
      blade. The blade's shaft moves along the black slit. Two rows of
      dyes have been injected into the surface of the fluid. The lower
      set is composed of a sequence of blue and red dyes, which have
      had an initial distance of $4.4$ cm from the blade's path. The upper
      blue dyes had an initial distance of $5$ cm from the
      slit. Snapshots have been taken when $\dot x>0$ and the blade is
      half way from the end points of its path. The stroke length is
      $AB=4.5$ cm. The resolution of camera is $1920\times 1080$
      pixels.  We have maximized the exposure with linear Gamma
      correction of the pictures to eliminate the color of the
      background fluid.}
		\label{fig2}
	}
\end{figure*}

Our experimental apparatus consists of a blade with two degrees of
freedom that moves in a viscous fluid. Figure \ref{fig1}C shows the
edge view of the blade, which undergoes a rectilinear motion in the
$x^{+}$ direction ($S^{+}$) with the constant angle of attack
$\theta$, then stops at point $B$ and rotates by $\pi-2\theta$ in the
clockwise direction ($R^{+}$). It then moves in the $x^{-}$ direction
($S^{-}$) until it stops at point $A$, makes a counter-clockwise turn
of $\pi-2\theta$ ($R^{-}$), and repeats the sequence
$S^{+} \rightarrow R^{+} \rightarrow S^{-} \rightarrow R^{-}$.
% It is to be noted that a simultaneous combination of linear
% rocking motion with rotational motion, i.e. the disk rotates in its
% rectilinear motion, results in a time symmetric motion that would
% not generate any motion at low Reynolds.
During the translational and rotational phases, the linear and angular
velocities are set to constant values $v_0$ and $\omega_0$. In
accordance with the gait of \textit{C. elegans}, we set
$\theta=45^{\circ}$. Sufficiently far from the blade where geometric
effects diminish, the velocity field becomes a Stokeslet
$\sim v_0 r^{-1}$ during $S^{\pm}$ stroke phases, and the rotations
$R^{\pm}$ generate rotlets scaled as $\sim \omega_0 r^{-2}$ with $r$
being the radial distance from the blade's center.

\section{Experimental Results}

We use a gantry $x$-$y$ table with a stepper motor M1 that controls the horizontal $x$-coordinate of a second servo motor M2. We couple motor M2 to a vertical shaft (in the $z$ direction) that holds a circular, aluminum blade at its other end. The blade rotates around the $z$-axis, and its angle of attack $\theta$ is controlled by motor M2. The diameter and thickness of the blade are $D=30\, {\rm mm}$ and $w=0.8\,{\rm mm}$, respectively. The fluid is corn syrup with density $\rho=1.34\, {\rm g/cm^3}$ and kinematic viscosity $\nu=41\, {\rm cm^2/s}$. In order to study the free surface flow, the blade is half submerged in the liquid, and we inject color dyes at the surface. We use red and blue food colors as tracer dyes and dissolve them in corn syrup before injecting into the surface. This minimizes the effect of diffusion and possible interactions between dyes \cite{cira2015vapour}. We introduce two rows of dyes on both sides of the line segment $AB$. A two-color sequence is used in the lower row to study and visualize the mixing function of the blade (Fig. \ref{fig2}).   

We actuate the blade by setting $v_0=2.5$ cm/s and $\omega_0=5\pi/2$
rad/s. These yield a low-Reynolds-number condition---with
$Re \approx 0.12$---relevant to the swimming and crawling of {\it
  C. elegans\/}. Each motion cycle has a period of $T=4\, {\rm
  s}$. Figure \ref{fig2} and Supplementary Material video 1
\cite{supp-videos} display
how initially round dyes evolve as the time elapses. They are first
pumped toward the blade, then stretched and folded as they cross the
$x$-axis. After passing through the blade, they form a wavelike
structure that propagates in the $y^{+}$ direction and is similar to
the trace that the head of a wild type {C. elegans\/} leaves behind
\cite{sznitman2010material}. The wavelength of the streaming field
decreases as tracers depart from the blade. This is due to the decline
in the velocity magnitude of the Stokeslet field. Furthermore,
  the rocking motion of blade creates circulatory streams on the two
  sides of the blade as it translates. To show this effect, we have
conducted an experiment by setting $\omega_0=0$ and traced the
trajectories of mm-size tracer particles using image processing
techniques. Supplementary Material video 2 \cite{supp-videos} shows how circulatory
streams form on the two sides of the blade.  The observed flow
structure
% (Fig. \ref{stream-line})
is reminiscent of recirculatory streams attached to each segment of
the body of {\it C. elegans\/}
\cite{gray1964locomotion,shen2011undulatory}, including its head.

The waving stream is then pulled sideways by rotlet fields generated
at points $A$ and $B$, folded as a kidney-shaped bundle, and sucked by
the blade from the lower side. The emerging circulatory flow diverts
fluid elements from the entire domain toward the line segment along
which the blade reciprocates. This is evident from the trajectories
that the stretched and folded blue dyes, initially located on the
upper side of the blade, follow until they reach the blade. The mixing
function of the mechanism is deduced from the disappearance of
initially blue and red colors and the emergence of a mixed color
within the kidney-shaped structure at $t=220\, {\rm s}$ and
afterwards. Mixing is characterized by the recurrence time of fluid
elements to the line segment $AB$. The larger the initial radial
distance from the blade, longer the circulation of deformed elements
and their revisit of the blade. Repeated stretching and folding of
fluid elements by $S^{\pm}$ Stokeslets, $R^{\pm}$ rotlets and the
blade, are analogous to the horseshoe map and suggest chaotic mixing
which we later numerically quantify.

% \begin{figure}[!h]
% \centerline{\mbox{\includegraphics[width=0.5\textwidth]{fig3.eps}}  }
% \caption{Forward pumping by the blade. (a) Fluid stream and its waveform with the decreasing wavelength $\lambda$. The temporal distances between successive peaks are identical. The front curve $F$ is expanding. The snapshot has been taken at $t=60$s. (b) Average velocity in the $y$ direction when the blade is passing through $x=0$. (c) Flow structure in the vicinity of the blade at $t=17$s. The region between the two dashed lines is impenetrable due to adhesion and surface tension effects.}
% \label{fig3}
% \end{figure}

\begin{figure}[!h]
  \parbox[c]{.48\linewidth}{\mbox{\includegraphics[width=0.5\textwidth]{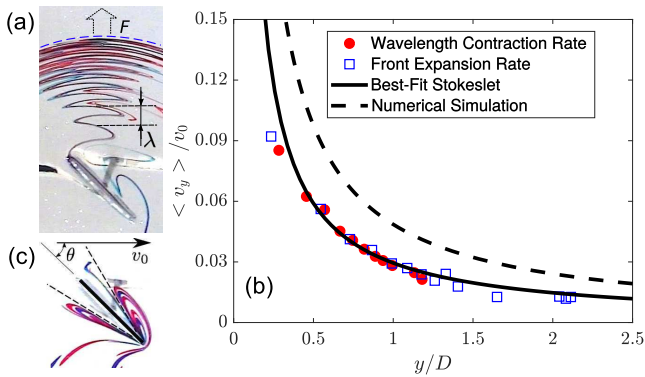}}}\hfill
  \parbox[c]{.50\linewidth}{
    \caption{Forward pumping by the blade. (a) The fluid stream and its
      waveform with the decreasing wavelength $\lambda$. The temporal
      distances between successive peaks are identical. The front
      curve $F$ is expanding. The snapshot has been taken at
      $t=60$s. (b) The average velocity in the $y$ direction when the
      blade is passing through $x=0$. (c) The flow structure in the
      vicinity of the blade at $t=17$s. The region between the two
      dashed lines is impenetrable due to adhesion and surface tension
      effects.}
\label{fig3}
  }
\end{figure}

The envelope of the kidney-shaped structure is expanding. This is an
indicator of the pumping function, which we quantify by measuring the
average streaming velocity $\langle v_y\rangle$ along the
$y$-axis. It is to be noted that in the rocking and rotating
  motion, the fluid is pushed in the normal direction to the blade's
  plane, and the averaged motion of fluid determines the net pumping
  effect and its direction. We measure the expansion rate through two
methods: the contraction rate of the wavelength $\lambda$, and the
expansion rate of the front $F$ (Fig. \ref{fig3}A). In the first
technique, we analyze the snapshot at $t=60\, {\rm s}$ well before the
mixing process dissolves the sharp features of the developing streamlines,
and in the second method, we follow the front curve $F$ in the strobe
images taken with increments of $T=4\, {\rm s}$. Figure \ref{fig3}B
illustrates the experimental values of $\langle v_y\rangle$ versus
$y$. Our mean experimental error level is about $4\%$, which is mainly
due to the perspective image distortions of the camera.

\section{Numerical Results}

We further test the setup using Finite Element Methods. We
  simplify the problem to a two dimensional flow where the fluid
  velocity $\mathbf{u}=(u,v)$ and pressure $p$ satisfy
\begin{eqnarray*}
    \Delta \mathbf{u} + \nabla p =0,\\    
    \nabla\cdot \mathbf{u} = 0.
\end{eqnarray*}
Each stage of blade's motion (i.e., $S^+, R^+, S^-, R^-$) is divided
into $N$ snapshots and velocity field is obtained using finite element
methods for each of the snapshots with proper boundary conditions
describing the blade's movement. The number $N$ is resolved until the
numerical result converges (here $N=100$). Open source Finite Element
Library \texttt{FreeFEM++} \cite{hecht2012new} is used to solve for
the weak form of the equation. The variables are non-dimensionalized
with $L=2$cm, $\mathcal{T}=1$s. The numerical domain is set to
$[0,9.5] \times [0,29]$ to match the experimental setup.  The
numerical simulation borders are seeded using nodes spaced by
$\delta x = 0.2$, and $\delta x = 0.005$ respectively; and the element
mesh is then created using the \texttt{FreeFEM++} mesh generator.

\begin{figure}[!h]
  \parbox[c]{.50\linewidth}{
    \mbox{\includegraphics[width=0.5\textwidth]{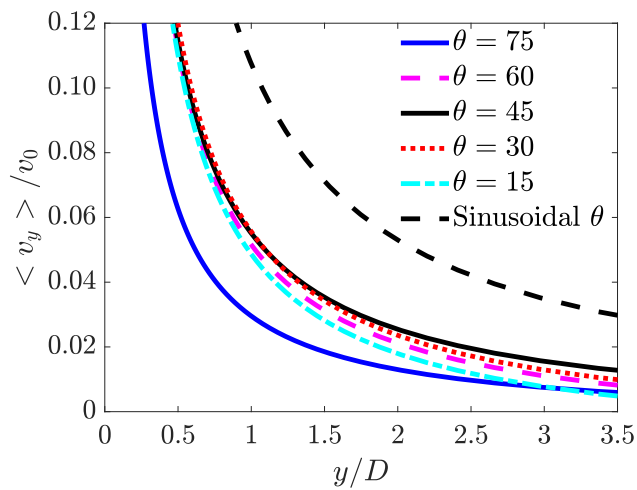}}}\hfill
  \parbox[c]{.5\linewidth}{
    \caption{Numerical simulation results for the average velocity of
      the wave front in the y direction when the blade is passing
      through x = 0 for different values of the disk angle
      i.e. $\theta$. The maximum velocity profile in the digitized
      motion is obtained at $\theta = 45$. A continuous sinusoidal
      rocking and rotational motion (dashed black line) has higher
      average velocity profile than the digitized versions of the motion. }
    \label{fig-phase-space}
}
\end{figure}

When the blade is at $x=0$, analytical calculations
\cite{jalali2015microswimmer,jalali2014versatile} give
$v_y/v_0=\Delta C\sin(2\theta)/[8\pi (y/D)]$ where $\Delta C$ is the
difference between the coefficients of drag forces against the broad
side and edge-wise motions of the blade. In the limit of a fully
  submerged, razor-thin disk ($w\rightarrow 0$) one finds
  $\Delta C=8/3$ \cite{happel2012low}.  Fitting an analytical
function $\sim (y/D)^{-1}$ to experimental data, gives
$\Delta C \approx 0.74$ (best-fit Stokeslet curve in
Fig. \ref{fig3}B). Alongside the analytical calculation, we computed
the front curve speed in our numerical simulation and found that it
follows a $(y/D)^{-1}$ trend close to the experiment's result
(numerical simulation curve in Fig. \ref{fig3}B).  To understand
the origin of the observed difference between experimental results and
analytical/numerical results, we have produced a close-up view
of the flow field around the blade at $t=17$s and displayed it in
Fig. \ref{fig3}C. Inspection of Fig. \ref{fig3}C shows that the flow
separates from the leading edge of the blade and an {\it impenetrable
  layer\/} develops between two dashed lines. The thickness of the
impenetrable layer depends on adhesion and surface tension effects
\cite{ren2009adhesion}. The blade and its associated
impenetrable layer behave like a moving wedge. Therefore, the
effective drag against edge-wise motion increases significantly and
reduces the front speed: instead of a simple skin drag on the two
faces of the disk, the edge-wise motion now experiences an extra {\it
  form drag\/}. Less adhesive surfaces can boost the pumping
efficiency. Since the flow is at low Reynolds ($Re \ll 1$), it is
  apparent that 
  % the blade's drag coefficient is linearly proportional to its
  % velocity. As a result
  increasing blade's velocity would effectively increase the pumping
  velocity. On the other hand, the net effect of blade's angle is less
  obvious; therefore we study the effect of blade angle in our
digitized motion (see Fig. \ref{fig-phase-space}). The front
speed first increases with the increase in blade's angle until a
maximum velocity profile is reached at $\theta = \pi/4$. After this
optimal value, the velocity profile decreases significantly with the
increase in blade's angle. Note that at $\theta = 0$ the rotational
motion is present and contributes to pumping and mixing; however, at
$\theta = \pi/2$ the rotational motion vanishes and the net motion
becomes a time reversal translation which is unable to generate any
pumping at low Reynolds. We further analyze a simultaneous
  sinusoidal rocking and rotational motion with
  $X = (l_s/2) \sin (2\pi t/T + \pi/2)$,
  $\theta = \pi/2 - \pi/4 \sin (2\pi t/T)$, where $l_s$ is the stroke
  length, and $T$ is the period of motion. As shown in
  Fig. \ref{fig-phase-space}, this simultaneous sinusoidal rocking and
  rotational motion creates a much higher wave front speed; however,
  because of experimental setup complexity is not studied here.

% This continuous sinusoidal motion however complicates our
% experimental setup and is not studied here.

% \begin{figure}[!h]
% \centerline{\mbox{\includegraphics[width=0.45\textwidth]{fig4.eps}}}
% \caption{Numerical simulation results of average velocity of wave
%   front in the y direction when the blade is passing through x = 0 for
%   different values of the disk angle i.e. $\theta$. The maximum
%     velocity profile in the digitized motion is obtained at
%     $\theta = 45$. A continuous sinusoidal rocking and rotational
%     motion (dashed black line) has higher average velocity profile
%     than digitized versions of motion. }
% \label{fig-phase-space}
% \end{figure}

% \begin{figure}[!h]
% \includegraphics[width=0.5\textwidth]{fig5.eps}
% \put(-58,177){$\mathcal{L}$}
% \put(-218,235){\textbf{(a)}}
% % \put(-98,95){\textbf{(b)}}
%  \put(-83,235){\textbf{(b)}}
%   \put(-89,73){$T$}
%   \put(-217,120){\textbf{(c)}}
%   \caption{(a) The profile of $\lambda_{\infty} (\mathcal{L}, 100T)$
%     along the line segment $\mathcal{L}$, (b) The state of the line
%     element represented by 200 tracer particles at $t=0$ (red dots)
%     and $t=100T$ (black dots). The line segment is at the location of
%     the lower set of dye in Fig. \ref{fig2}. The position of the disks
%     are shown in dark and light blue. (c) Time evolution of the
%       maximum eigen-value over time for the mid-point $x=0$ in line
%       segment $\mathcal{L}$. The time period of motion is $T$ and the
%       inset figure shows a close up of the box shown. The time
%       evolution of maximum eigen-value is a continuous curve with
%       kicks that correspond to $R^+$ and $R^-$ segments of motion. }
% \label{fig_lyapunov}
% \end{figure}

\begin{figure}
  \parbox[c]{.50\linewidth}{
    \includegraphics[width=0.5\textwidth]{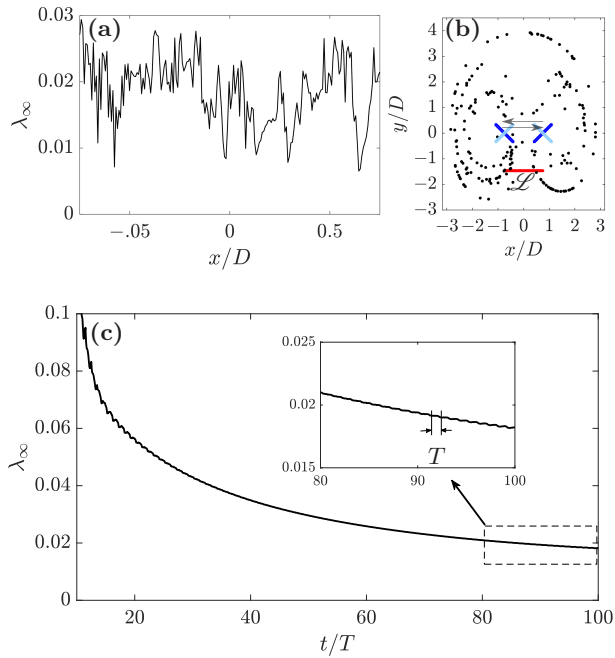}
    \put(-58,177){$\mathcal{L}$} \put(-218,235){\textbf{(a)}}
    % \put(-98,95){\textbf{(b)}}
    \put(-83,235){\textbf{(b)}} \put(-89,73){$T$}
    \put(-217,120){\textbf{(c)}}}
  \hfill \parbox[c]{.49\linewidth}{
    \caption{(a) The profile of $\lambda_{\infty} (\mathcal{L}, 100T)$
    along the line segment $\mathcal{L}$, (b) The state of the line
    element represented by 200 tracer particles at $t=0$ (red dots)
    and $t=100T$ (black dots). The line segment is at the location of
    the lower set of dye in Fig. \ref{fig2}. The positions of the disks
    are shown in dark and light blue. (c) Time evolution of the
      maximum eigen-value over time for the mid-point $x=0$ in line
      segment $\mathcal{L}$. The time period of motion is $T$ and the
      inset figure shows a close up of the box shown. The time
      evolution of maximum eigen-value is a continuous curve with
      kicks that correspond to $R^+$ and $R^-$ segments of motion. }
\label{fig_lyapunov}
  }
\end{figure}

% \begin{figure}[!h]
% \includegraphics[width=0.5\textwidth]{fig5ab.eps}
% \put(-48,37){$\mathcal{L}$}
% %\put(-245,95){\textbf{(a)}}
% \put(-222,100){\textbf{(a)}}
% % \put(-98,95){\textbf{(b)}}
%  \put(-75,100){\textbf{(b)}}\\
%   \includegraphics[width=.5\textwidth]{fig5c.eps}
%   \put(-89,73){$T$}
%   \put(-217,117){\textbf{(c)}}
%   \caption{(a) The profile of $\lambda_{\infty} (\mathcal{L}, 100T)$
%     along the line segment $\mathcal{L}$, (b) The state of the line
%     element represented by 200 tracer particles at $t=0$ (red dots)
%     and $t=100T$ (black dots). The line segment is at the location of
%     the lower set of dye in Fig. \ref{fig2}. The position of the disks
%     are shown in dark and light blue. (c) Time evolution of the
%       maximum eigen-value over time for the mid-point $x=0$ in line
%       segment $\mathcal{L}$. The time period of motion is $T$ and the
%       inset figure shows a close up of the box shown. The time
%       evolution of maximum eigen-value is a continuous curve with
%       kicks that correspond to $R^+$ and $R^-$ segments of motion. }
% \label{fig_lyapunov}
% \end{figure}

\begin{figure}[!h]
  \parbox[c]{.25\linewidth}{
    \mbox{\includegraphics[width=0.25\textwidth]{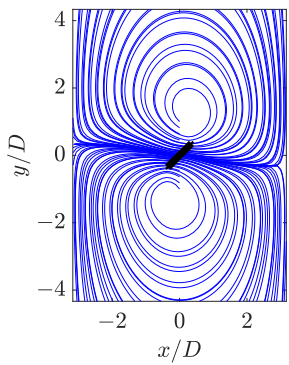}}}\hfill
  \parbox[c]{.75\linewidth}{
    \caption{Flow-field streamlines computed from instantaneous
      velocity field in the vicinity of the blade (shown with black
      lines) when the blade is passing through $x=0$. The motion
      corresponds to rectilinear $S^+$ motion. Due to symmetry,
      streamlines for $S^-$ motion are the mirror image of $S^{+}$
      streamlines with respect to $x=0$.}
    \label{stream-line}
}
\end{figure}

We utilize the finite time Lyapunov exponent to quantify the
  mixing process \cite{lapeyre2002characterization}. We
  calculate finite time largest Lyapunov exponent
  $\lambda_{\infty}$ along a line segment corresponding to the red dye
  shown in Fig. \ref{fig2} i.e
  $\mathcal{L}= \{(x,y)| -l_s/2 \leq x \leq l_s/2, y/D = -1.45 \}$
  where $l_s$ is the length of stroke. A positive Lyapunov exponent,
  $\lambda_{\infty}>0$, implies chaos and therefore an exponential
  divergence of neighboring traces lying on $\mathcal{L}$. To compute
  the Lyapunov exponent for a particle initially at $X_0$, we
  integrate
\begin{align}
  & \frac{\text{d}}{\text{d} t} M(t)  = \mathbf{J}(\mathbf{X}(t),t) ^{\top} M(t), \nonumber \\
  &  \quad \mathbf{J} = \nabla \mathbf{u}, \quad M(0) = \mathbb{I},\nonumber 
\end{align}
where $M$ is called resolvent matrix. The finite time Lyapunov
exponent $\lambda_{\infty}$ is then obtained as the logarithm of
maximum eigenvalue of $(M^{\top}M)^{1/2t}$. Figure \ref{fig_lyapunov}
shows the variation of $\lambda_{\infty}(\mathcal{L}, t = 100T)$. The
state of 200 particles along $\mathcal{L}$ after $t=100T$ is shown in
Fig. \ref{fig_lyapunov}B. The time evolution of the maximum
  eigen-value for a sample point, the mid-point particle with $x=0$,
  is shown in Fig. \ref{fig_lyapunov}C. The time evolution of maximum
  eigenvalue is a continuous line with kicks that corresponds to the
  $R^+$ and $R^-$ section of motion. We would like to point out that
  the Finite Element Simulation is a two dimensional low Reynolds
  approximation of the flow around the blade and it does not account
  for the free surface in the experimental setup. The free surface
  introduces surface tension at the blade's point of contact which
  further reduces the pumping efficiency.
  % and also (ii)
  % assumes a two dimensional low Reynolds flow, while in the
  % experiments the circular blade creates a three dimensional
  % flow. The
  % two dimensional approximation is valid as long as blade's border
  % free surface in the experiments introduces surface tension at the
  % blade which reduces the pumping efficiency
  % We would like
  % to point out that the Finite Element Simulation (i) does not
  % account
  % for the free surface that appears in the experiment, and also (ii)
  % assumes a two dimensional low Reynolds flow, while in the
  % experiments the circular blade creates a three dimensional
  % flow. The
  % two dimensional approximation is valid as long as blade's border
  % free surface in the experiments introduces surface tension at the
  % blade which reduces the pumping efficiency,
 Lastly, to show that the observed flow structure is reminiscent of
recirculatory streams attached to each segment of the body of {\it
  C. elegans\/} \cite{gray1964locomotion,shen2011undulatory}, the
numerically obtained circulatory motion of flow-field as the blade
passes through $x=0$ is shown in Fig. \ref{stream-line}.

% \begin{figure}[!h]
%   \centerline{\mbox{\includegraphics[width=0.25\textwidth]{fig6.eps}}}
%   \caption{Flow-field streamlines computed from instantaneous velocity
%     field in the vicinity of the blade (shown with black lines) when
%     the blade is passing through $x=0$. The motion corresponds to
%     rectilinear $S^+$ motion. Due to symmetry, streamlines for $S^-$ motion are
%     the mirror image of $S^{+}$ streamlines with respect to $x=0$.}
% \label{stream-line}
% \end{figure}

\begin{figure*}
\centerline{
\fbox{\includegraphics[width=0.95\textwidth]{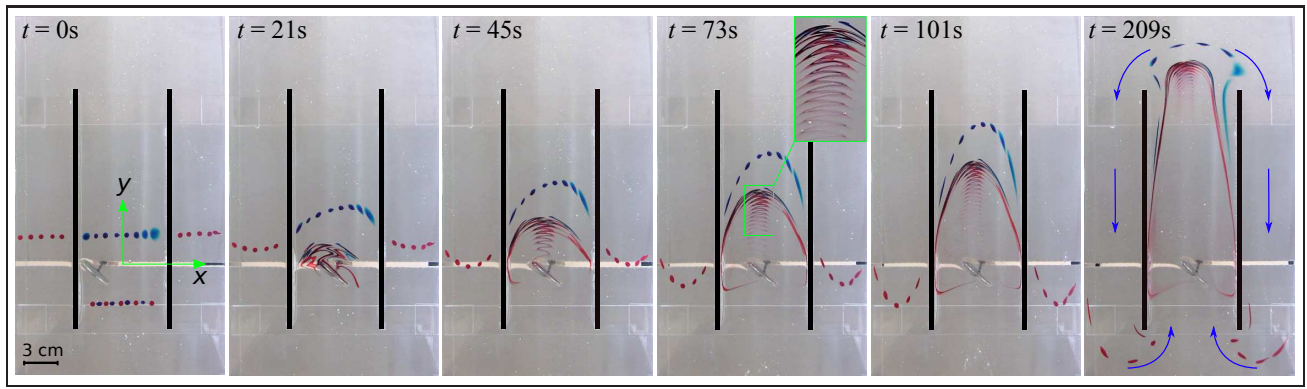}} 
}
\caption{Pumping between parallel walls of distance $77$ mm. The blade reciprocates along the $x$-axis, and makes $90^{\circ}$ rotations near the walls. The blade's angle of attack is $45^{\circ}$. The flow circulates from the side channels and returns to the central channel where the blade is moving.}
\label{fig4}
\end{figure*}

\begin{figure}[!h]
  \parbox[c]{.48\linewidth}{
    \mbox{\includegraphics[width=0.45\textwidth]{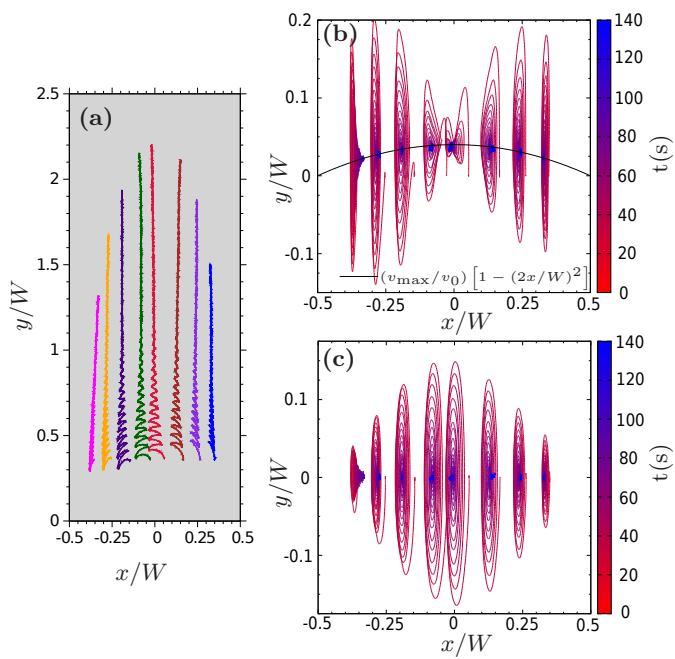}
      \put(2,57){\rotatebox{90}{t(s)}}
      \put(2,174){\rotatebox{90}{t(s)}} \put(-214,192){\textbf{(a)}}
      \put(-122,224){\textbf{(b)}} \put(-122,101){\textbf{(c)}}
      \put(-240,114){\rotatebox{90}{$y/W$}} \put(-200,20){$x/W$}
      \put(-77,-7){$x/W$} \put(-77,115){$x/W$} \put(-100,133){\tiny
        $(v_{\text{max}}/v_0)\left[ 1-(2x/W)^2\right] $}
      \put(-142,47){\rotatebox{90}{$y/W$}}
      \put(-142,162){\rotatebox{90}{$y/W$}} }}\hfill
  \parbox[c]{.48\linewidth}{
    \caption{Image processing results for eight tracer particles. (a)
      Trajectories of floated polystyrene spheres in the central
      channel (different colors represent different spheres). They
      start their motions from the line $y=0.35 W$. The blade
      reciprocates along the $x$-axis ($y=0$), and starts its motion
      from the leftmost endpoint. The initial zigzagging motion of tracers
      is due to wall effect, which diminishes at a distance $y\sim W$
      from the blade's location. (b) Evolution of the $v_y$ component
      of the velocity vector for eight tracer particles. Variable line
      colors indicate the elapsed time in seconds. It is seen that the
      velocities of particles along the central channel converge to
      Poiseuille profile with $v_{y}/v_0 = 0.04 [1-(2x/W)^2]$ (note
      blue regions). The velocity oscillation is due to the zigzagging
      motion. (c) Same as panel b but for $v_x$, which gradually tends
      to zero for all particles.}
\label{Fig-S1}
}
\end{figure}

\section{The Pumping Function}

We now operate the blade against two parallel separators of distance
$W=77$ mm. Figure \ref{fig4} shows the arrangement of the separators
in the main rectangular container, and the location of the blade. Four
sets of dyes have been introduced to the surface: two sets in side
channels between the separators and container walls, and two rows in
the main central channel on the two sides of the blade's path (see
Fig. \ref{fig4}, $t=0$). The blade is half-submerged and induces a
surface flow. Since the central channel is connected to the side
channels, we have a closed circuit for a possible free-surface
flow. We actuate the blade with the same conditions as in our previous
experiment. Figure \ref{fig4} and Supplementary Material video 3
\cite{supp-videos} show flow
generation along the central and side channels. In contrast to the
blade in infinite medium, where the streaming velocity drops as
$\sim r^{-1}$, we find that the fluid velocity reaches a steady
pattern along the channel. The blade sucks the fluid from the back
side and pumps it forward. The zigzag footprint of the blade's motion
carried by traveling dyes (inset figure at $t=73$s) is similar to the
head/tail-tip trajectory of wild-type \textit{C. elegans\/} (see
Fig. 1a in \cite{sznitman2010material}). Mixing is achieved over
circulation time scale as fluid elements repeatedly revisit the
blade. It is to be noted that the flow structure of a fixed blade
imparting a net force is fundamentally different to that of a force
free swimmer and here the the blade's motion is inspired by the head
movements of \textit{C. elegans.}

We have floated polystyrene beads on the fluid surface and used their
trajectories to measure the streaming velocity
(Fig. \ref{Fig-S1}). The velocity profile is Poiseuille and its
maximum occurs along the centerline of the middle channel. We find
$v_{\rm max}/v_0 \approx 0.04$. The trajectories of the polystyrene
tracers oscillate when they are close to the blade, but evolve into
straight lines at a distance $y \sim W$. Initial zigzagging movements
are due to wall effects: the moving blade induces two Stokeslet
components, one parallel to the wall and the other normal to it. The
normal component generates two side vortices \cite{lauga2005brownian} which trap
tracers until the parallel component pushes them out of the
wall-induced vortices. The swimming efficiency of
\textit{C. elegans\/} is defined as $U/c$, where $U$ is the swimming
speed and $c$ is the speed of bending waves. Similarly, the pumping
efficiency in our mechanism can be defined as $v_{\rm
  max}/v_0$. Depending on its environment, \textit{C. elegans\/} has a
swimming efficiency of $0.08 < U/c < 0.15$ \cite{shen2011undulatory}. The
  pumping efficiency in our experiments is about $4\%$. Therefore, a
  single blade's pumping efficiency is quite comparable with the
  worm's swimming efficiency, noting the fact that {\it C. elegans\/}
  gains thrust force from its tail, body and head segments ($U/c$
  comes from at least three segments and not just from the head), and
  that $\Delta C$ in our experiments is relatively low because of
  adhesion effects.

A simple way to enhance the pumping efficiency is to increase the
parameter $\Delta C$ by minimizing the adhesion between the fluid and
the blade. In the ideal case with no-slip boundary condition at the
blade's surface, and when the blade is fully submerged, one obtains
$\Delta C= 8/3$ and the efficiency is expected to reach its maximum
possible level of $0.14$.

We have not investigated the sensitivity of $v_{\rm max}/v_0$ to
variations of $v_0$, $\omega_0$ and $W/D$ because of two limiting
aspects of our setup: (i) decreasing $W/D$ connects the impenetrable
layers of the blade and the wall due to adhesion effects and halts
pumping (ii) increasing $v_0$ or $\omega_0$ generates large-amplitude
surface waves and influences the dynamics of tracers. The blade cannot
pump (as {\it C. elegans\/} cannot swim in highly viscous fluids) if
either of its reciprocating or rotational motions is disabled. A
flapping flexible blade, without translational motion, may still pump
but the efficiency will be extremely low, and it will not have mixing
capability. In sub-millimeter scales, the blade can be fabricated from
magnetic materials and actuated remotely by an external, oscillating
magnetic field. An implanted blade in an organ-on-a-chip device can
perform the pumping function of the heart while simultaneously mixing
and homogenizing the blood. In microprocessor technology, the blade
can perform the role of a heat exchanger unit by pumping polymeric
cooling liquid into a chamber and mixing heated and cooled streams as
they reach the blade.
%%%%%%%%%%%%%%%%%%%%%%%%%%%%%%%%%%%%%%%%%%%%%%%%%%%%%%%%%%%%%%%%%%%%%%%%%%%%5

\section{Conclusion}
  
  In this manuscript, we created a low Reynold artificial mechanism
  with the goal of combining mixing and pumping functions into a
  single device, while each function requires a separate unit in the
  existing microfluidic devices. Inspired by the motion of {\it
    C. elegans}, which are evolved swimmers in highly viscous
  environments, we analyzed the motion of the head segment by
  approximating the head with a disk and breaking the motion into two
  stages of rocking and rotational movement. We showed that the
  blade's motion not only mixes the surrounding fluid in a chaotic
  way, but also generates a steady Poiseuille flow with a pumping
  efficiency comparable with the swimming efficiency of the worm. We
  further showed that maximum efficiency actually occurs for the head
  motion of {\it C. elegans}, i.e., when the maximum disk angle is
  $45^o$ and the rocking and rotational stages follow a sinusoidal
  function with a $\pi/2$ phase difference. The pumping and mixing
  function resulted by the blade motion have potential applications in
  microfluidic devices such as microprocessors cooled by polymetric
  flows.

%%%%%%%%%%%%%%%%%%%%%%%%%%%%%%%%%%%%%%%%%%%%%%%%%%%%%%%%%%%%%%%%%%%%%%%%%%%%5
\section{ACKNOWLEDGMENTS} 
This project was supported by the National Science Foundation under
grant No.  CMMI-1562871. The authors thank Leo Brossollet and Robert
McKnight for their help in the initial development of the experimental
setup.  We would also like to thank the anonymous referees for
valuable comments.

%
%In our experiment, camera is attached below
%the corn syrup container. The color dye used in the experiments is
%much darker than the corn syrup (i.e. background liquid). Therefore,
%it is possible to locate the dye and syrup frontier based on the
%contrast in their color. Initially, we convert the recorded video
%stream into a sequence of RGB frames. Then, each colored image is
%converted to a grayscale format, and finally into a binary image
%(i.e. black and white) via an appropriate threshold value (see Fig. 1
%for more details). With an appropriate threshold value, the background
%liquid becomes white while the dye mixture appear as black region. To
%quantize the propulsion (or pumping) speed of the dye wave, the linear
%motion of a point on dye frontier has been recorded. 
%
%This point is the
%intersection of the perpendicular bisector of AB and the front wave
%(\ref{}Fig. 1). Note that A and B are two end points of the blade stroke
%where it changes its direction. Although this point has an oscillating back and forth motion
%in short time scales (due to up and down motion of the blade), its
%average velocity is upward and increasing.}

\bibliography{ref.bib}
%%%%%%%%%%%%%%%%%%%%%%%%%%%%%%%%%%%%%%%%%%%%%%%%%%%%%%%%%%%%

%%%%%%%%%%%%%%%%%%%%%%%%%%%%%%%%%%%%%%%

\section{Appendix} 
\appendix

\section*{Image processing method}

\begin{figure}
  % \centerline{\mbox{\includegraphics[width=0.45\textwidth]{fig1.eps}}  }
  \includegraphics[width=1.\textwidth]{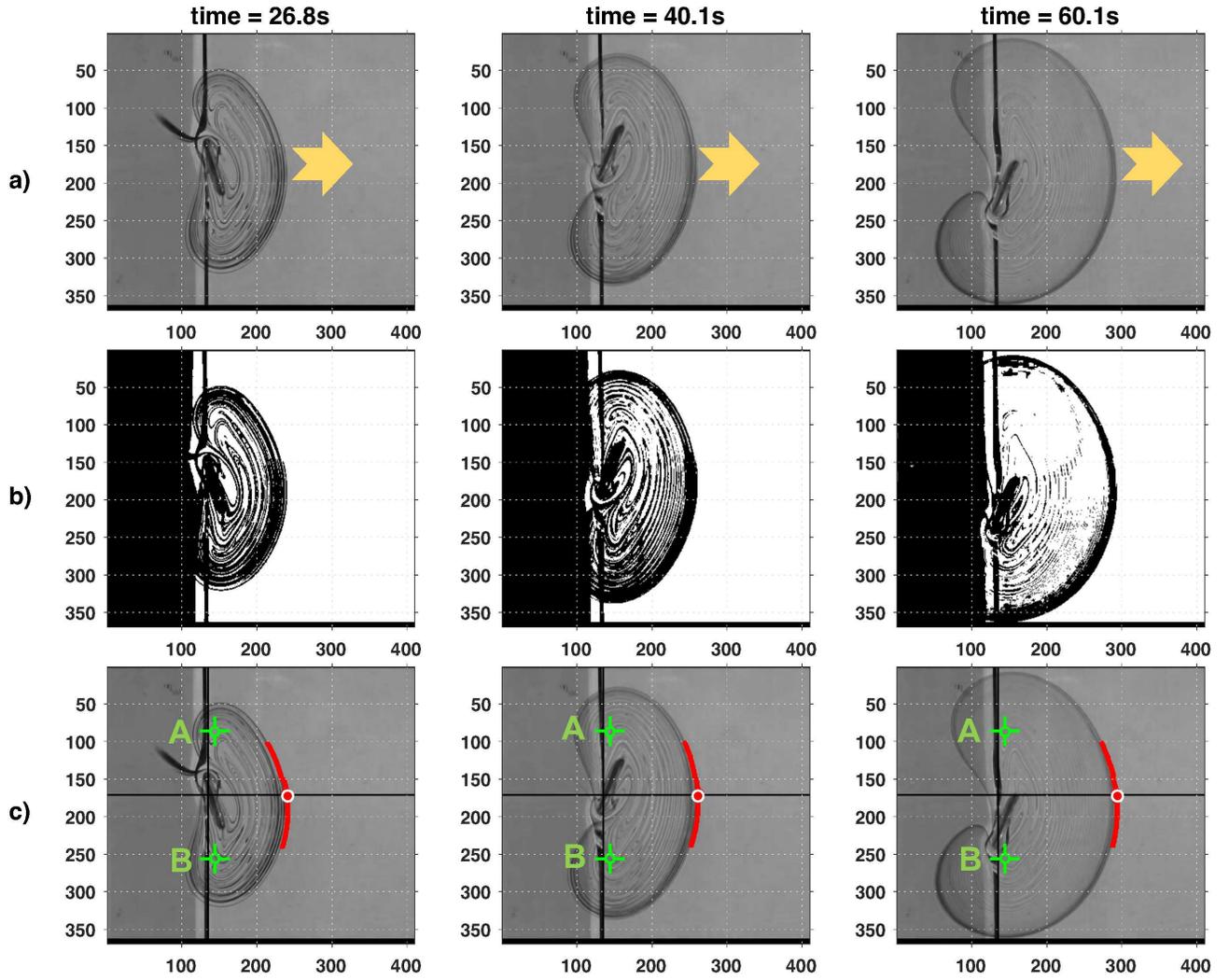}
  \caption{ Three different snapshots show how the initial column of
    dye propagates by the blade's motion. (a) Image extracted from the
    recorded video stream; (b) binary (i.e. black and white)
    conversion of the RGB image reveals the frontier of the dye wave;
    (c) edge of dye mixture is identified based on color contrast
    between the background liquid (corn syrup) and color dye.}
\label{supp-fig1}
\end{figure}

A camera is installed under the corn syrup container facing
upward. Red and blue food color dyes are injected in the liquid at its
surface in order to track and quantize the liquid motion. Since the
color dyes used in the experiment are much darker than the background
liquid (i.e. corn syrup), wave frontier line is detected using color
contrast. After converting the recorded video stream into a sequence of
RGB frames, each color frame is transformed into a grayscale and then
a binary image based on an appropriate threshold value to
differentiate the dye frontier from the background liquid (the
frontier appears in black while the background liquid turns into white
in the binary image; see Fig. \ref{supp-fig1}). To quantize the
propulsion/pumping speed of the dye wave, the linear motion of the
frontier wave center point is calculated and tracked. This point is
the intersection of the perpendicular bisector of AB and the front
wave (see Fig. 2 in the manuscript). While this point has an
oscillating back and forth motion in short time scales due to rocking
motion of the blade, its cycle average speed is always to the right
(Fig. \ref{supp-fig1}).

\section*{Effect of Blade Angle and Rocking Velocity on Pumping (or
  Propulsion) Velocity}

To investigate the effect of blade angle and rocking velocity on
pumping performance, a set of experiments with different combination
of blade angle and velocity is conducted. The results of two rocking
velocities are shown in Fig. \ref{fig2} and \ref{fig3}. For a given
blade angle and rocking velocity, the pumping speed at a given
location is a function of distance from the blade (denoted by $X$). As
the front boarder moves away from the blade, its linear expansion rate
decreases. Additionally, for a constant blade rocking velocity, the
maximum expansion rate occurs at an optimal blade angle
$\theta \approx 45^\circ$. As the rocking speed increases, the flow
diverges from low Reynolds flow and, the maximum expansion rate
happens at larger angles i.e. $\theta \approx 50^{\circ}$ (More
details can be found in Fig. \ref{fig3})

\begin{figure}
  % \centerline{\mbox{\includegraphics[width=0.45\textwidth]{fig1.eps}}  }
  \includegraphics[width=0.90\textwidth]{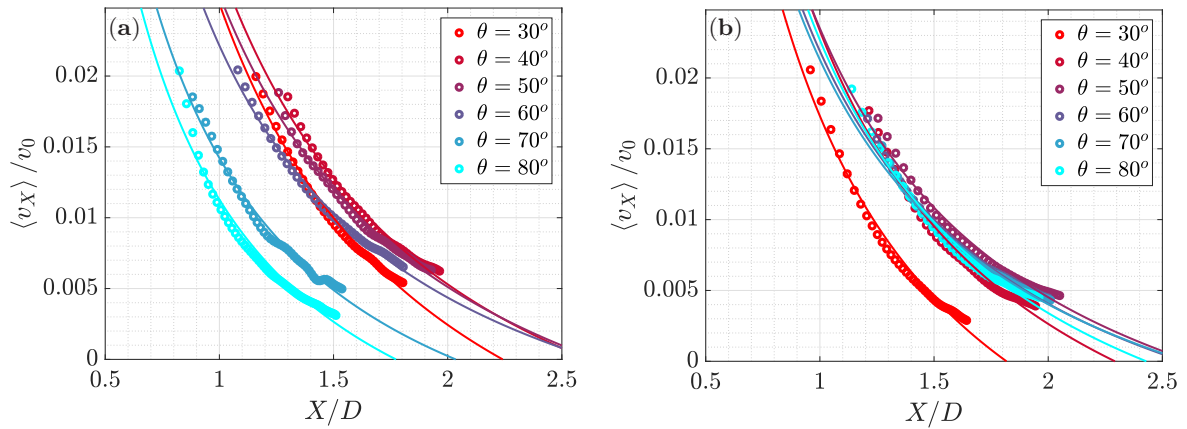}
  \put(-421,150){$\mathbf{(a)}$}
  \put(-194,150){$\mathbf{(b)}$}
  \caption{Effect of blade angle $\theta$ and rocking velocity $v_0$
    on pumping speed. Different blade angles $\theta$ are tested for
    rocking speed (a) $v_{0}=4.87$cm/s; (b) $v_0=7.3$ cm/s. $D$ is the
    blade's diameter.}
\label{fig2}
\end{figure}

\begin{figure}
  % \centerline{\mbox{\includegraphics[width=0.45\textwidth]{fig1.eps}}  }
  \includegraphics[width=0.90\textwidth]{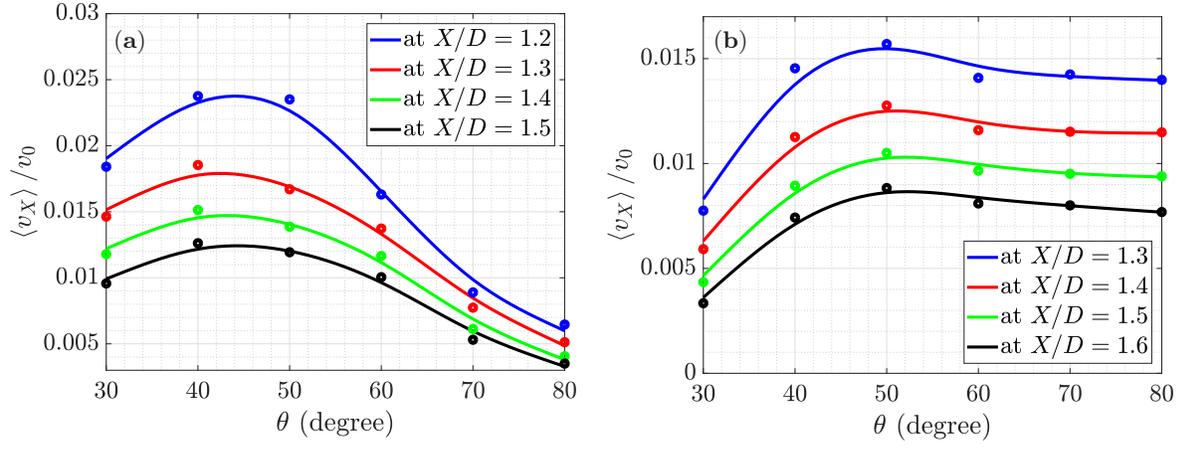}
  \put(-419,150){$\mathbf{(a)}$}
  \put(-192,150){$\mathbf{(b)}$}
  \caption{Maximum pumping speed at a given location $X/D$ versus
    blade angle $\theta$ for rocking velocity (a) $v_0=4.87$ cm/s, and
    (b) $v_0=7.3$ cm/s.  }
\label{fig3}
\end{figure}

\end{document}